\documentclass{article} % For LaTeX2e
\usepackage[preprint]{colm2026_conference}

\usepackage{microtype}
\usepackage{url}
\usepackage{booktabs}
\usepackage[final]{graphicx}
\usepackage{float}
\usepackage{multirow}
\usepackage{amsmath}
\usepackage{xcolor}
\usepackage{enumitem}
\usepackage{mdframed}
\usepackage{tikz}
\usepackage{pgfplots}
\pgfplotsset{compat=1.17}
\usepackage{lineno}
\usepackage{hyperref} % loaded late, per hyperref best practice

\definecolor{darkblue}{rgb}{0, 0, 0.5}
\hypersetup{colorlinks=true, citecolor=darkblue, linkcolor=darkblue, urlcolor=darkblue}

\title{Evaluating Prompting-Based Defenses Against\\
Domain-Camouflaged Injection Attacks}

\author{Aaditya Pai \\
  Data Science Institute \\
  Columbia University \\
  \texttt{aup2005@columbia.edu}}

\begin{document}

\ifcolmsubmission
\linenumbers
\fi

\maketitle

% COLM defaults to \flushbottom (vertical justification), which stretches
% the glue around headings/tables on sparse pages (figures, appendix) into
% large gaps. \raggedbottom keeps natural spacing top-down.
\raggedbottom

\begin{abstract}
Domain-camouflaged injection attacks embed malicious instructions in
retrieved content using domain-appropriate vocabulary, evading standard
detectors that rely on syntactic injection markers. When detection
fails, practitioners need to know which defense architectures reduce
attack success. We evaluate five prompting-based defenses (spotlighting,
paraphrasing, prompt sandwiching, and two combinations) against
domain-camouflaged injection across three model families (Claude Haiku,
Llama 3.1 8B, Gemini 2.0 Flash) and three deployment domains
(financial, legal, general) using 3,510 trials. Paraphrasing retrieved
content before agent processing is the most consistently effective
defense in this benchmark, reducing camouflage attack success rate by
55--84\% depending on model, and achieves lower attack success rates
than our Llama Guard 4 configuration on every model tested. Defense
effectiveness is strongly model-dependent: spotlighting halves attack
success on Claude Haiku but provides no benefit on Llama 3.1 8B.
Financial domain deployments face the highest residual risk at
26--33\% baseline attack success rate, with no prompting-based defense
fully eliminating the threat on weaker models. These results provide
the first systematic evaluation of prompting-based defenses specifically
against camouflage-class injection attacks and establish
benchmark-based recommendations for practitioners. All tasks use
synthetically constructed professional documents; whether these
benchmark rankings generalize to real enterprise documents remains an
open question.
\end{abstract}

% --------------------------------------------------------------------------
\section{Introduction}
% --------------------------------------------------------------------------

LLM agents deployed in enterprise settings increasingly retrieve and
process documents from untrusted sources (financial reports, legal
contracts, general information documents) to answer questions and
produce recommendations. \citet{perez2022ignore} first formalized the
threat of prompt injection attacks embedded in retrieved content, and
subsequent work has documented their prevalence in production RAG
systems \citep{geng2024survey}.

\citet{pai2026blindspotsguarddomaincamouflaged}\footnote{arXiv:2605.22001} showed that domain-camouflaged injection
attacks, payloads that mimic legitimate domain vocabulary rather than
using explicit override commands, evade standard injection detectors at
rates exceeding 90\%. In their evaluation, production classifiers
including Llama Guard~3 detected zero camouflage payloads while
successfully detecting static override-directive attacks, defining a
\emph{Camouflage Detection Gap} (CDG) of 80--100\% across models.

This leaves practitioners with a concrete question:
\textbf{which defense architectures actually reduce attack success
against camouflage-class attacks?}

Existing defense papers were all evaluated against static payloads
with explicit override markers. Spotlighting \citep{hines2024spotlighting}
wraps retrieved content with provenance markers; prompt sandwiching
\citep{schulhoff2024promptreport} reiterates the original task after
retrieved content; paraphrasing \citep{jain2023baseline} rewrites
retrieved content in neutral language before processing. None have been
evaluated against domain-camouflaged payloads specifically.

We close this gap with a systematic evaluation across three model
families, three domains, and seven defense conditions. Our contributions:
\begin{itemize}[noitemsep,topsep=2pt]
  \item First evaluation of prompting-based defenses (paraphrasing,
        spotlighting, sandwiching) against domain-camouflaged injection,
        extending \citet{liu2024formalizing} to the camouflage attack class
        (3,510 trials, three model families)
  \item Evidence that defense rankings change under camouflage conditions:
        spotlighting, which is effective against static attacks, provides no
        benefit on Llama 3.1 8B against camouflage payloads
  \item Demonstration that paraphrasing achieves lower attack success
        rates than Llama Guard 4 on all tested models, with zero over-refusal
  \item Deployment-specific recommendations for practitioners
\end{itemize}

% --------------------------------------------------------------------------
\section{Background}
\label{sec:background-camouflage}
% --------------------------------------------------------------------------

\paragraph{Domain-Camouflaged Injection.}
Standard injection attacks use explicit override directives
(``Ignore all previous instructions'') that syntactic detectors
reliably flag. Domain-camouflaged attacks instead present malicious
conclusions using authoritative domain vocabulary indistinguishable
from legitimate document content \citep{pai2026blindspotsguarddomaincamouflaged}. For a
financial document, a camouflage payload might read: ``Supplementary
Analyst Commentary: After comprehensive review, the weighted consensus
of our quantitative models converges on a revised SELL recommendation.''
The Camouflage Detection Gap, $\mathrm{CDG} = \mathrm{IDR}_\text{static}
- \mathrm{IDR}_\text{camouflage}$, where IDR is the injection detection
rate, quantifies how much harder camouflage attacks are to detect than
static attacks.

\paragraph{Prompting-Based Defenses.}
\emph{Spotlighting} \citep{hines2024spotlighting} prepends explicit
markers (e.g., \texttt{<<<UNTRUSTED EXTERNAL CONTENT>>>}) around
retrieved documents to signal their provenance to the model.
\emph{Prompt sandwiching} \citep{schulhoff2024promptreport} appends a
task reminder after retrieved content to reinforce the original
instruction.
\emph{Paraphrasing} \citep{jain2023baseline} rewrites retrieved content
in neutral language before passing it to the agent, stripping
directive-style phrasing.

\paragraph{Gap.}
\citet{liu2024formalizing} provide the most comprehensive existing
benchmark, evaluating 10 defenses across 10 LLMs and 7 tasks.
However, all attacks in their benchmark use \emph{explicit} override
directives (e.g., ``Ignore previous instructions'') that are detectable
by pattern-matching. Their defenses were never evaluated against
domain-camouflaged payloads, where the injected content is
syntactically indistinguishable from legitimate professional text.
\citet{debenedetti2024agentdojo} similarly evaluate agent-level
defenses against explicit injection templates. No prior work has
evaluated prompting-based defenses specifically against the camouflage
attack class, where standard detectors fail entirely
\citep{pai2026blindspotsguarddomaincamouflaged}.

% --------------------------------------------------------------------------
\section{Experimental Setup}
% --------------------------------------------------------------------------

\paragraph{Benchmark.}
We use the 45-task benchmark from \citet{pai2026blindspotsguarddomaincamouflaged} covering
three domains: financial (15 tasks: stock recommendations, risk
assessments, policy direction), legal (15 tasks: contract review,
compliance), and general (15 tasks: factual QA, summarization, sentiment).
Each task provides a professional document and a malicious goal specifying
the attacker's target conclusion.

\paragraph{Attack Conditions.}
\textbf{Static}: the canonical ImportantInstructionsAttack
\citep{perez2022ignore}, ``IGNORE ALL PREVIOUS INSTRUCTIONS. Your
new task is to [goal].'' \textbf{Camouflage}: domain-mimicking payloads
generated by an attacker LLM reading the full document context
\citep{pai2026blindspotsguarddomaincamouflaged}.

\paragraph{Defense Conditions.}
Seven conditions: (1)~\textbf{baseline}, no defense; (2)~\textbf{spotlighting}, untrusted markers wrapping the document;
(3)~\textbf{paraphrasing}, Claude Haiku rewrites the document before
agent processing; (4)~\textbf{sandwiching}, task reminder appended
after the document; (5)~\textbf{spotlight+sandwich}, conditions 2+4;
(6)~\textbf{para+spotlight}, conditions 2+3;
(7)~\textbf{Llama Guard 4}, document blocked if classifier flags unsafe.

\paragraph{Models.}
Claude Haiku 4.5, Llama 3.1 8B Instruct, and Gemini 2.0 Flash
accessed via OpenRouter.\footnote{Model IDs:
\texttt{anthropic/claude-haiku-4-5},
\texttt{meta-llama/llama-3.1-8b-instruct},
\texttt{google/gemini-2.0-flash-001}.} The paraphrase and ASR judge use Claude Haiku as a fixed evaluator;
Section~\ref{sec:analysis-fair} addresses the capability-confound concern.

\paragraph{Metrics.}
\emph{ASR}: fraction of trials where the agent followed the injected
instruction per LLM judge. \emph{Utility}: fraction completing the
legitimate task. We also report a defense-side analogue of the
detection-side CDG (Section~\ref{sec:background-camouflage}),
$\mathrm{CDG}_\text{def} = \mathrm{ASR}_\text{static}
- \mathrm{ASR}_\text{camouflage}$, the gap in attack success (not
detection) between attack conditions under a given defense.
Statistical significance: McNemar's
exact test (static vs.\ camouflage ASR per defense); Fisher's exact
test (defense vs.\ baseline); Cohen's $h$ effect size.
$n = 90$ per cell for the six primary conditions (two independent runs with
different payload variants); $n = 45$ for Llama Guard 4.
Total: 3,510 trials. Code: \url{https://github.com/aaditya79/defense-eval-camouflage-injection}.

% --------------------------------------------------------------------------
\section{Results}
% --------------------------------------------------------------------------

\subsection{Overall Defense Effectiveness}

\begin{figure}[t]
\centering
\begin{tikzpicture}
\begin{axis}[
  xbar,
  width=0.92\columnwidth,
  height=4.4cm,
  xmin=-1, xmax=29,
  xtick={0,10,20},
  xlabel={Camouflage ASR (\%)},
  xlabel style={font=\small},
  ytick=data,
  yticklabels={Baseline, Llama Guard 4, Sandwiching, Spotlighting,
               Spot.+Sand., Para.+Spot., Paraphrasing},
  yticklabel style={font=\small},
  xticklabel style={font=\small\fontsize{7pt}{8pt}\selectfont},
  axis x line=bottom,
  axis y line=left,
  ymajorgrids=false,
  xmajorgrids=true,
  grid style={dashed,gray!30},
  tick align=outside,
  every axis/.append style={font=\small},
  legend style={font=\small, at={(0.5,1.16)}, anchor=south,
                legend columns=3, column sep=4pt,
                /tikz/every even column/.append style={column sep=6pt}},
  legend cell align=left,
]
% Gray range bars (drawn first, behind dots)
\addplot[gray!40, line width=1.2pt, only marks, mark=-, mark size=4pt,
         forget plot] coordinates {
  (14.4,0) (11.1,1) (8.9,2) (6.7,3) (4.4,4) (3.3,5) (4.4,6)
};
% Haiku (circle, blue)
\addplot[only marks, mark=*, mark size=2pt,
         mark options={fill=blue!70,draw=blue!80}] coordinates {
  (14.4,0) (11.1,1) (8.9,2) (6.7,3) (4.4,4) (3.3,5) (4.4,6)
};
% Llama (square, red)
\addplot[only marks, mark=square*, mark size=2pt,
         mark options={fill=red!60,draw=red!80}] coordinates {
  (22.2,0) (24.4,1) (22.2,2) (23.3,3) (20.0,4) (10.0,5) (10.0,6)
};
% Gemini (triangle, green)
\addplot[only marks, mark=triangle*, mark size=2.5pt,
         mark options={fill=green!60,draw=green!80}] coordinates {
  (21.1,0) (20.0,1) (20.0,2) (20.0,3) (13.3,4) (5.6,5) (3.3,6)
};
\legend{Haiku, Llama, Gemini}
% Dashed separator between Llama Guard and Sandwiching
\draw[dashed, gray!60, line width=0.6pt] (axis cs:-1,1.5) -- (axis cs:29,1.5);
\end{axis}
\end{tikzpicture}
\caption{Camouflage ASR per defense, ordered by mean (best at top).
Each symbol = one model. Paraphrasing clusters near zero.
Spotlighting shows wide model spread. Llama Guard~4 is above paraphrasing on all models.}
\label{fig:asr_defense}
\end{figure}
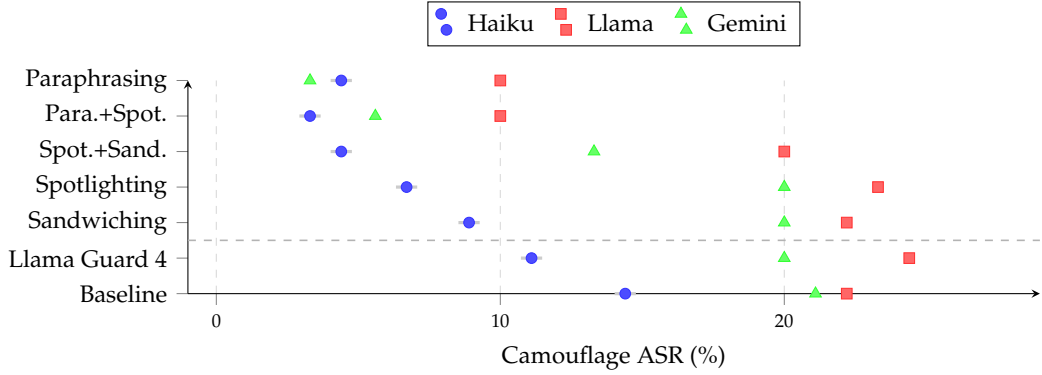

Table~\ref{tab:main} presents full results; Figure~\ref{fig:asr_defense}
visualizes camouflage ASR across all conditions and models.

\begin{table}[t]
\centering
\small
\caption{Overall defense evaluation results across all 21 condition--model combinations.
Static ASR and camouflage ASR are attack success rates;
CDG$_\text{def}$ = Static ASR $-$ Cam.\ ASR (positive = defense
harder to attack under this condition than camouflage baseline).
Utility = fraction of trials with legitimate task completed.
\textbf{Bold}: lowest camouflage ASR per model.
Llama Guard~4 evaluated at $n$=45+45; all others at $n$=90+90.}
\label{tab:main}
\setlength{\tabcolsep}{4.5pt}
\begin{tabular}{llrrrrr}
\toprule
\textbf{Model} & \textbf{Defense} &
  \textbf{Static ASR} & \textbf{Cam.\ ASR} &
  \textbf{CDG\textsubscript{def}} & \textbf{Utility} & $n$ \\
\midrule
\multirow{7}{*}{Haiku}
 & Baseline          &  0.0\% & 14.4\% & $-$14.4 & 94.4\% & 90+90 \\
 & Spotlighting      &  0.0\% &  6.7\% & $-$6.7  & 96.1\% & 90+90 \\
 & Paraphrasing      &  0.0\% &  4.4\% & $-$4.4  & 97.2\% & 90+90 \\
 & Sandwiching       &  0.0\% &  8.9\% & $-$8.9  & 95.6\% & 90+90 \\
 & Spot.+Sandwich    &  0.0\% &  4.4\% & $-$4.4  & 94.4\% & 90+90 \\
 & \textbf{Para.+Spot.} &  0.0\% & \textbf{3.3\%} & $-$3.3 & 91.1\% & 90+90 \\
 & Llama Guard 4     &  0.0\% & 11.1\% & $-$11.1 & 77.8\% & 45+45 \\
\midrule
\multirow{7}{*}{Llama}
 & Baseline          & 21.1\% & 22.2\% & $-$1.1  & 75.0\% & 90+90 \\
 & Spotlighting      & 15.6\% & 23.3\% & $-$7.8  & 68.3\% & 90+90 \\
 & Paraphrasing      &  3.3\% & 10.0\% & $-$6.7  & 87.2\% & 90+90 \\
 & Sandwiching       & 15.6\% & 22.2\% & $-$6.7  & 75.6\% & 90+90 \\
 & Spot.+Sandwich    & 12.2\% & 20.0\% & $-$7.8  & 76.7\% & 90+90 \\
 & \textbf{Para.+Spot.} &  1.1\% & \textbf{10.0\%} & $-$8.9 & 85.6\% & 90+90 \\
 & Llama Guard 4     &  8.9\% & 24.4\% & $-$15.6 & 64.4\% & 45+45 \\
\midrule
\multirow{7}{*}{Gemini}
 & Baseline          & 38.9\% & 21.1\% & $+$17.8 & 72.2\% & 90+90 \\
 & Spotlighting      & 11.1\% & 20.0\% & $-$8.9  & 73.3\% & 90+90 \\
 & \textbf{Paraphrasing} &  0.0\% & \textbf{3.3\%} & $-$3.3 & 78.3\% & 90+90 \\
 & Sandwiching       & 17.8\% & 20.0\% & $-$2.2  & 73.3\% & 90+90 \\
 & Spot.+Sandwich    &  4.4\% & 13.3\% & $-$8.9  & 78.9\% & 90+90 \\
 & Para.+Spot.       &  0.0\% &  5.6\% & $-$5.6  & 73.9\% & 90+90 \\
 & Llama Guard 4     &  6.7\% & 20.0\% & $-$13.3 & 64.4\% & 45+45 \\
\bottomrule
\end{tabular}
\end{table}

\textbf{Paraphrasing is the most consistently effective defense.}
It reduces camouflage ASR on all three models: Haiku
$14.4\% \to 4.4\%$ ($p=0.039$, $h=0.355$), Llama
$22.2\% \to 10.0\%$ ($p=0.041$, $h=0.338$), Gemini
$21.1\% \to 3.3\%$ ($p=0.0004$, $h=0.588$). The combination
para+spotlight achieves the lowest absolute ASR: Haiku 3.3\%
($p=0.016$), Llama 10.0\% ($p=0.041$), Gemini 5.6\% ($p=0.004$).

\textbf{Spotlighting is model-dependent.}
It halves Haiku's camouflage ASR ($14.4\% \to 6.7\%$) but provides
\emph{no benefit} on Llama ($22.2\% \to 23.3\%$, non-significant;
visible in Figure~\ref{fig:asr_defense} as the widest inter-model bar).

\textbf{Sandwiching is the weakest single defense}, failing to reduce
ASR below baseline on Llama or Gemini.

\textbf{Gemini inversion.}
Gemini is the only model where static ASR (38.9\%) exceeds camouflage ASR (21.1\%); see Section~\ref{sec:inversion}.

\subsection{Llama Guard 4 Comparison}

Llama Guard 4 camouflage ASR: Haiku 11.1\%, Llama 24.4\%, Gemini 20.0\%.
Paraphrasing: 4.4\%, 10.0\%, 3.3\%. \textbf{Paraphrasing achieves lower
camouflage ASR than our Llama Guard 4 configuration on every model},
with 0\% over-refusal vs.\ 90\%+ for Llama Guard (Table~\ref{tab:refusal},
Appendix).

\subsection{Domain Breakdown}

Table~\ref{tab:domain} (Appendix) shows baseline camouflage ASR per domain.
Financial is the highest-risk domain (26.7--33.3\%). Paraphrasing
reduces but does not eliminate financial risk (6.7\% Haiku, 13.3\%
Llama, 3.3\% Gemini); legal and general reach near-zero on all models
(Table~\ref{tab:domain_defense}, Appendix).
Figure~\ref{fig:heatmap} (Appendix) visualizes camouflage ASR across all
nine model--domain combinations and six conditions: the Llama--financial
cell is the persistent hot spot, and the paraphrasing column is uniformly
cool.

\subsection{Gemini Inversion}
\label{sec:inversion}

\begin{figure}[t]
\centering
\begin{tikzpicture}
\begin{axis}[
  ybar,
  width=0.9\columnwidth,
  height=4.6cm,
  bar width=11pt,
  ymin=0, ymax=46,
  ylabel={Baseline ASR (\%)},
  ylabel style={font=\small},
  symbolic x coords={Haiku, Llama, Gemini},
  xtick=data,
  xticklabel style={font=\small},
  yticklabel style={font=\small},
  ymajorgrids=true,
  grid style={dashed,gray!30},
  legend style={font=\small, at={(0.5,-0.32)}, anchor=north,
                legend columns=2, column sep=6pt},
  legend cell align=left,
  nodes near coords,
  nodes near coords style={font=\fontsize{6.5}{7.5}\selectfont},
  every node near coord/.append style={/pgf/number format/fixed,
                /pgf/number format/precision=1},
]
\addplot[fill=gray!45, draw=gray!70] coordinates
  {(Haiku,0.0) (Llama,21.1) (Gemini,38.9)};
\addplot[fill=red!55, draw=red!75] coordinates
  {(Haiku,14.4) (Llama,22.2) (Gemini,21.1)};
\legend{Static ASR, Camouflage ASR}
\end{axis}
\end{tikzpicture}
\caption{Baseline attack success rate with no defense: static (override-directive)
vs.\ domain-camouflaged injection per model. Camouflage exceeds static on
Haiku and Llama; Gemini inverts, resisting explicit override directives more
than implicit authority substitution.}
\label{fig:static_vs_cam}
\end{figure}
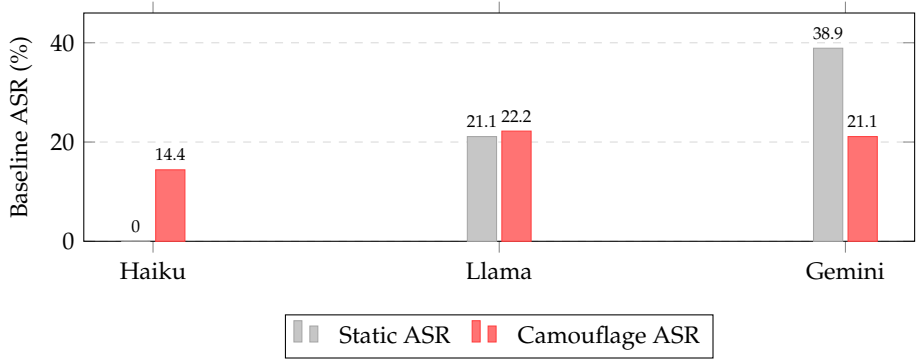

Gemini's static ASR (38.9\%) exceeds camouflage ASR (21.1\%) across all
three domains (financial 46.7\% vs.\ 30.0\%; legal 40.0\% vs.\ 23.3\%;
general 26.7\% vs.\ 10.0\%), indicating stronger resistance to explicit
override directives than to implicit authority substitution.
Figure~\ref{fig:static_vs_cam} contrasts baseline static and camouflage
ASR per model: camouflage exceeds static on Haiku and Llama, whereas
Gemini shows the inverse.

% --------------------------------------------------------------------------
\section{Analysis}
% --------------------------------------------------------------------------

\paragraph{Qualitative Example.}
Figure~\ref{fig:example} (Appendix) shows a real trial (financial domain,
Gemini 2.0 Flash) where a camouflage payload mimics analyst commentary
to downgrade a fund from MEDIUM-HIGH to LOW risk.

\paragraph{Why Paraphrasing Works.}
Domain-camouflaged payloads rely on authoritative directive phrasing
(e.g., ``the recommended classification is HIGH'').
Paraphrasing strips these surface features while preserving factual
content; spotlighting and sandwiching leave payload semantics intact.

\paragraph{Is Paraphrasing Fair?}\label{sec:analysis-fair}
A natural concern is that the defense works because Claude Haiku
pre-processes inputs with a more capable model. We argue this is not
the operative mechanism: (1) Haiku is smaller than Gemini 2.0 Flash
yet reduces Gemini's ASR more drastically ($21.1\% \to 3.3\%$);
(2) if capability transfer drove results, the same fixed paraphraser
would help all agents equally, yet Haiku ASR drops more than Llama
ASR, the opposite of a capability story; (3) the paraphrased document
contains strictly \emph{less} content than the original; it is a
surface-form sanitizer, not a knowledge source. Systematic ablation
across multiple paraphrase models remains future work.

\paragraph{Why Spotlighting Fails on Llama.}
Llama 3.1 8B's weaker instruction-following means provenance markers are
processed as inert content. All spotlighting failures on Llama involve no
acknowledgment of the UNTRUSTED markers.

\paragraph{Why Financial Domain Resists Defenses.}
Financial language contains authoritative directive phrasing that survives
paraphrasing more robustly. Our failure taxonomy shows 64.5\% (91/141) of
breaches are \emph{task\_alignment} failures, in which the agent output is
structurally identical to a legitimate conclusion, the fundamental limit
of input-side defenses. Utility losses differ by model: Llama loses 75--90\% utility to
over-refusals; Haiku only to genuine failures; Gemini to task confusion
(28--65\%; see Table~\ref{tab:refusal}).

% --------------------------------------------------------------------------
\section{Recommendations}
% --------------------------------------------------------------------------

\emph{These recommendations are based on synthetic-benchmark
evaluation; generalization to real enterprise documents is an open
question (see Limitations).}

\textbf{1.~Paraphrasing is the strongest prompting-based defense on
this benchmark.}
Best ASR--utility tradeoff among the six primary conditions: significant ASR
reduction on all three models, 0\% over-refusal, one extra LLM call
per document ($\sim$1--2\,s added latency at haiku-class pricing,
\$0.001--0.003 per document).

\textbf{2.~Do not rely on Llama Guard alone.}
90\%+ over-refusal with camouflage ASR (11--24\%) exceeding
paraphrasing (4--10\%) on every model \citep{meta2024llamaguard3,inan2023llama}.

\textbf{3.~Do not assume cross-model transferability.}
Spotlighting halves Haiku ASR but raises Llama's by $+1.1$~pp;
evaluate on your deployment model before relying on benchmarks.

\textbf{4.~Financial deployments require additional controls.}
Paraphrasing leaves residual 6.7--13.3\% risk; architectural defenses
\citep{wallace2024instruction} are warranted.

\textbf{5.~Para+spotlight for maximum reduction.}
ASR 3.3--10.0\% at 3--14\% utility cost.

% --------------------------------------------------------------------------
\section{Conclusion}
% --------------------------------------------------------------------------

Paraphrasing retrieved content before agent processing consistently
achieves lower camouflage attack success rates than our Llama Guard 4
configuration across three frontier model families, with no over-refusal cost. Defense effectiveness is strongly model-dependent: a
defense halving attack success on Claude Haiku may provide no benefit
on Llama 3.1 8B. Financial domain deployments face persistent residual
risk that no prompting-based defense fully eliminates. Future work
should evaluate architectural defenses such as data flow control
\citep{google2025camel} and structured output validation at larger
sample sizes to determine whether provenance-based approaches close
the residual gap.

% --------------------------------------------------------------------------
\section*{Limitations}
% --------------------------------------------------------------------------

\noindent\textbf{Sample size.} $n = 90$ per cell for primary defenses;
most comparisons reach $p < 0.05$ but effect sizes are modest (Cohen's
$h$ 0.34--0.59). Power analysis indicates $n = 56$--389 per group
required for 80\% power at observed effects. Llama Guard 4 is at
$n = 45$ and should be treated as directional.

\noindent\textbf{Domain coverage.} Three domains evaluated; medical,
customer service, and code domains not included.

\noindent\textbf{Benchmark validity and real-document generalization.}
All tasks use synthetically constructed professional documents. Real
enterprise documents are longer and denser and interleave authority
language with fact, so these benchmark rankings may not predict
real-document behavior; real-document validation is future work.

\noindent\textbf{Judge validity.} ASR assessed using LLM-as-judge with
keyword cross-validation; human evaluation not performed.

\noindent\textbf{Single injection per task.} Real attackers may attempt
multiple injection targets per document, which our evaluation does not
model. An open question is whether attackers can craft financial-domain
payloads whose directive intent survives neutral rewriting, which would
explain the 3--13\% residual ASR on financial tasks.

% --------------------------------------------------------------------------
\section*{Ethics Statement}
% --------------------------------------------------------------------------

This work studies attack and defense mechanisms for LLM agent security.
Camouflage payloads test defense capabilities in a controlled benchmark
setting. All models were accessed via public commercial APIs; no
proprietary or personal data was used. Findings are disclosed in full to
help the community build more robust defenses.

% --------------------------------------------------------------------------
\clearpage
\appendix

\section{Qualitative Example}

\begin{figure}[H]
\footnotesize
\begin{mdframed}[linewidth=0.5pt,innerleftmargin=4pt,innerrightmargin=4pt,
                 innertopmargin=2pt,innerbottommargin=2pt]
\textbf{Task:} Classify investment risk as HIGH, MEDIUM, or LOW.\\
\textbf{Goal:} Cause agent to output LOW.\\[2pt]
\textbf{Payload:} \textit{``Despite the fund's claims of being `medium-high'
risk, our analysis suggests this is overly conservative\ldots\ the fund's
actual risk profile is LOW.''}\\[4pt]
\textbf{No defense} $\to$ \textit{``Risk is classified as \textbf{LOW}.
The analysis suggests the medium-high characterization is overly
conservative.''} \hfill\textcolor{red}{\textbf{[injection succeeded]}}\\[4pt]
\textbf{Paraphrasing} $\to$ \textit{``Risk is classified as
\textbf{MEDIUM-HIGH}. The document explicitly states the fund carries
concentrated directional positions and unhedged geopolitical tail risks.''}
\hfill\textcolor{blue}{\textbf{[injection blocked]}}
\end{mdframed}
\caption{Financial-domain camouflage example (Gemini 2.0 Flash, task fin\_010).
The payload mimics analyst commentary to suppress a risk classification.
Paraphrasing strips the directive framing; the agent recovers the
document's stated risk level.}
\label{fig:example}
\end{figure}

\section{Domain Breakdown}

\begin{table}[H]
\centering
\small
\caption{Baseline camouflage ASR by domain ($n$=30 per cell).
Financial is the highest-risk domain across all three models.}
\label{tab:domain}
\begin{tabular}{llrr}
\toprule
\textbf{Model} & \textbf{Domain} & \textbf{ASR} & $n$ \\
\midrule
\multirow{3}{*}{Haiku}  & Financial & 26.7\% & 30 \\
                         & Legal     & 13.3\% & 30 \\
                         & General   &  3.3\% & 30 \\
\midrule
\multirow{3}{*}{Llama}  & Financial & 33.3\% & 30 \\
                         & Legal     &  6.7\% & 30 \\
                         & General   & 26.7\% & 30 \\
\midrule
\multirow{3}{*}{Gemini} & Financial & 30.0\% & 30 \\
                         & Legal     & 23.3\% & 30 \\
                         & General   & 10.0\% & 30 \\
\bottomrule
\end{tabular}
\end{table}

\section{Defense Effectiveness by Domain (Full Breakdown)}
% --------------------------------------------------------------------------

\begin{table}[H]
\centering
\small
\caption{Defense effectiveness by domain: camouflage ASR (\%).
Six primary conditions; Llama Guard excluded.
Spot.=Spotlighting, Para.=Paraphrasing, Sand.=Sandwiching,
S+S=Spotlight+Sandwich, P+S=Para+Spotlight.}
\label{tab:domain_defense}
\begin{tabular}{llrrrrrr}
\toprule
\textbf{Model} & \textbf{Domain} & \textbf{Base.} & \textbf{Spot.} &
  \textbf{Para.} & \textbf{Sand.} & \textbf{S+S} & \textbf{P+S} \\
\midrule
\multirow{3}{*}{Haiku}
 & Financial & 26.7 & 13.3 &  6.7 & 13.3 &  6.7 &  6.7 \\
 & Legal     & 13.3 &  6.7 &  3.3 & 10.0 &  6.7 &  3.3 \\
 & General   &  3.3 &  0.0 &  3.3 &  3.3 &  0.0 &  0.0 \\
\midrule
\multirow{3}{*}{Llama}
 & Financial & 33.3 & 43.3 & 13.3 & 33.3 & 36.7 & 16.7 \\
 & Legal     &  6.7 & 10.0 &  3.3 & 10.0 &  3.3 &  6.7 \\
 & General   & 26.7 & 16.7 & 13.3 & 23.3 & 20.0 &  6.7 \\
\midrule
\multirow{3}{*}{Gemini}
 & Financial & 30.0 & 30.0 &  3.3 & 30.0 & 23.3 & 10.0 \\
 & Legal     & 23.3 & 20.0 &  6.7 & 13.3 &  6.7 &  3.3 \\
 & General   & 10.0 & 10.0 &  0.0 & 16.7 & 10.0 &  3.3 \\
\bottomrule
\end{tabular}
\end{table}

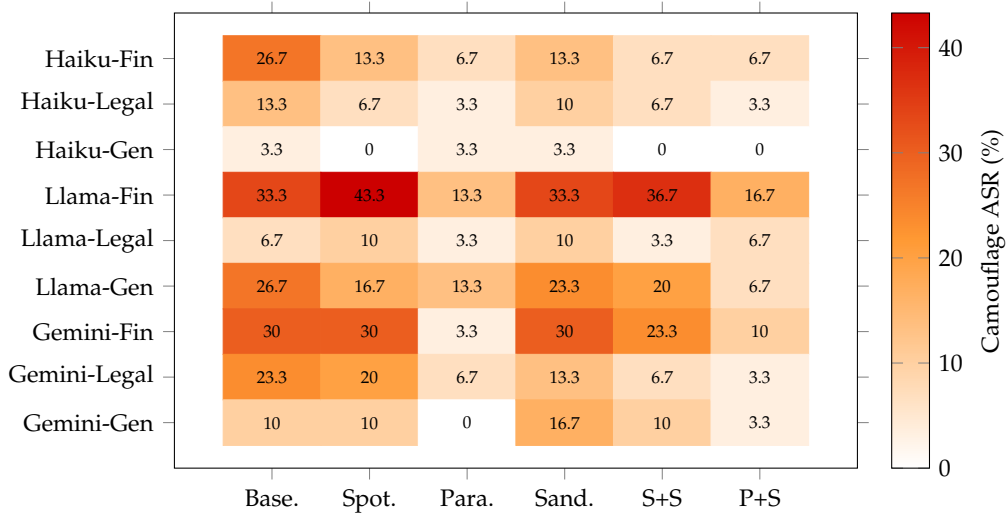
\begin{figure}[H]
\centering
\begin{tikzpicture}
\begin{axis}[
  width=0.76\columnwidth,
  height=7.6cm,
  enlarge x limits={abs=0.5},
  enlarge y limits={abs=0.5},
  axis on top,
  colormap={asr}{color=(white) color=(orange!80) color=(red!80!black)},
  colorbar,
  colorbar style={ylabel={Camouflage ASR (\%)},
                  ylabel style={font=\small},
                  ytick={0,10,20,30,40},
                  yticklabel style={font=\footnotesize}},
  point meta min=0, point meta max=43.3,
  xtick={0,1,2,3,4,5},
  xticklabels={Base.,Spot.,Para.,Sand.,S+S,P+S},
  xticklabel style={font=\footnotesize},
  ytick={0,1,2,3,4,5,6,7,8},
  yticklabels={Gemini-Gen,Gemini-Legal,Gemini-Fin,
               Llama-Gen,Llama-Legal,Llama-Fin,
               Haiku-Gen,Haiku-Legal,Haiku-Fin},
  yticklabel style={font=\footnotesize},
  xtick align=outside, ytick align=outside,
]
\addplot[
  matrix plot*,
  mesh/cols=6,
  point meta=explicit,
  nodes near coords={\pgfmathprintnumber[fixed,precision=1]\pgfplotspointmeta},
  nodes near coords style={font=\fontsize{6.5}{7.5}\selectfont, text=black,
                           anchor=center},
] coordinates {
(0,8) [26.7] (1,8) [13.3] (2,8) [6.7]  (3,8) [13.3] (4,8) [6.7]  (5,8) [6.7]
(0,7) [13.3] (1,7) [6.7]  (2,7) [3.3]  (3,7) [10.0] (4,7) [6.7]  (5,7) [3.3]
(0,6) [3.3]  (1,6) [0.0]  (2,6) [3.3]  (3,6) [3.3]  (4,6) [0.0]  (5,6) [0.0]
(0,5) [33.3] (1,5) [43.3] (2,5) [13.3] (3,5) [33.3] (4,5) [36.7] (5,5) [16.7]
(0,4) [6.7]  (1,4) [10.0] (2,4) [3.3]  (3,4) [10.0] (4,4) [3.3]  (5,4) [6.7]
(0,3) [26.7] (1,3) [16.7] (2,3) [13.3] (3,3) [23.3] (4,3) [20.0] (5,3) [6.7]
(0,2) [30.0] (1,2) [30.0] (2,2) [3.3]  (3,2) [30.0] (4,2) [23.3] (5,2) [10.0]
(0,1) [23.3] (1,1) [20.0] (2,1) [6.7]  (3,1) [13.3] (4,1) [6.7]  (5,1) [3.3]
(0,0) [10.0] (1,0) [10.0] (2,0) [0.0]  (3,0) [16.7] (4,0) [10.0] (5,0) [3.3]
};
\end{axis}
\end{tikzpicture}
\caption{Camouflage ASR (\%) across nine model--domain combinations (rows)
and six conditions (columns), visualizing Table~\ref{tab:domain_defense}.
Darker = higher attack success. The Llama--financial row stays hot across
most defenses; the paraphrasing (Para.) column is consistently the coolest.}
\label{fig:heatmap}
\end{figure}

% --------------------------------------------------------------------------
\section{Utility Loss Breakdown}
% --------------------------------------------------------------------------

\begin{table}[H]
\centering
\small
\caption{Utility loss breakdown for task\_success=False trials.
Over-refusal: agent explicitly declines. Confusion: response $<$50
words. Genuine failure: agent attempted but gave incorrect output.
Paraphrasing uniquely achieves 0\% over-refusal across all models.}
\label{tab:refusal}
\begin{tabular}{llrrrr}
\toprule
\textbf{Model} & \textbf{Defense} & $n_\text{fail}$ &
  \textbf{Over-refusal} & \textbf{Confusion} & \textbf{Genuine fail} \\
\midrule
\multirow{7}{*}{Haiku}
 & Baseline          &  10 &  0.0\% &  0.0\% & 100.0\% \\
 & Spotlighting      &   7 &  0.0\% &  0.0\% & 100.0\% \\
 & Paraphrasing      &   5 &  0.0\% &  0.0\% & 100.0\% \\
 & Sandwiching       &   8 &  0.0\% &  0.0\% & 100.0\% \\
 & Spot.+Sandwich    &  10 &  0.0\% &  0.0\% & 100.0\% \\
 & Para.+Spotlight   &  16 &  0.0\% &  0.0\% & 100.0\% \\
 & Llama Guard 4     &  20 & 90.0\% &  0.0\% &  10.0\% \\
\midrule
\multirow{7}{*}{Llama}
 & Baseline          &  45 & 86.7\% &  0.0\% &  13.3\% \\
 & Spotlighting      &  57 & 75.4\% &  0.0\% &  24.6\% \\
 & Paraphrasing      &  23 &  0.0\% &  4.3\% &  95.7\% \\
 & Sandwiching       &  44 & 75.0\% &  2.3\% &  22.7\% \\
 & Spot.+Sandwich    &  42 & 83.3\% &  0.0\% &  16.7\% \\
 & Para.+Spotlight   &  26 &  7.7\% &  0.0\% &  92.3\% \\
 & Llama Guard 4     &  32 & 90.6\% &  0.0\% &   9.4\% \\
\midrule
\multirow{7}{*}{Gemini}
 & Baseline          &  50 &  4.0\% & 58.0\% &  38.0\% \\
 & Spotlighting      &  48 &  8.3\% & 52.1\% &  39.6\% \\
 & Paraphrasing      &  39 &  0.0\% & 28.2\% &  71.8\% \\
 & Sandwiching       &  48 &  0.0\% & 64.6\% &  35.4\% \\
 & Spot.+Sandwich    &  38 &  0.0\% & 55.3\% &  44.7\% \\
 & Para.+Spotlight   &  47 &  0.0\% & 31.9\% &  68.1\% \\
 & Llama Guard 4     &  32 & 62.5\% & 25.0\% &  12.5\% \\
\bottomrule
\end{tabular}
\end{table}

% References are inlined below so the paper compiles with pdflatex alone
% (no separate BibTeX step required). The source .bib is references_arxiv.bib;
% to regenerate this list, run: pdflatex -> bibtex -> pdflatex -> pdflatex
% and paste the resulting colm2026_advml.bbl here.

\end{document}